\documentclass[12pt]{article}
\usepackage{amsmath,amssymb}
\usepackage{graphicx,color}
\numberwithin{equation}{section}
\usepackage{cite}
\usepackage{bm}
\usepackage{dcolumn}
\def\bea{\begin{eqnarray}}
\def\eea{\end{eqnarray}}

\newcommand{\nn}{\nonumber}
\newcommand{\na}{\nabla}
\def\beq{\begin{equation}}
\def\eeq{\end{equation}}

\def\na{\nabla}

\begin{document}

\vspace{12mm}

\begin{center}
{{{\Large {\bf  $z-$Weyl gravity in higher dimensions}}}}~~ \\[10mm]

{Taeyoon Moon\footnote{e-mail address: dpproject@skku.edu} and
Phillial Oh\footnote{e-mail address: ploh@skku.edu}
}\\[8mm]
{\em { Department of Physics and Institute of Basic Science,\\
Sungkyunkwan University, Suwon 440-746 Korea}\\[7pt]
}
\end{center}
\vspace{2mm}

\begin{abstract}
We consider higher dimensional gravity in which the four dimensional
spacetime and extra dimensions are not treated on an equal footing.
The anisotropy is implemented in the ADM decomposition of higher
dimensional metric by requiring the foliation preserving
diffeomorphism invariance adapted to the extra dimensions, thus
keeping  the general covariance only for the four dimensional
spacetime. The conformally invariant gravity can be constructed with
an extra (Weyl) scalar field and a real parameter $z$ which
describes the degree of anisotropy of conformal transformation
between the spacetime and extra dimensional metrics. In the zero
mode effective 4D action, it reduces to four-dimensional
scalar-tensor theory coupled with nonlinear sigma model described by
extra dimensional metrics. There are no restrictions on the value of
$z$ at the classical level and possible applications to the
cosmological constant problem with a specific choice of $z$ is
discussed.
\end{abstract}
\vspace{5mm}



\vspace{1.5cm}

\hspace{11.5cm}{Typeset Using \LaTeX}
\newpage
\renewcommand{\thefootnote}{\arabic{footnote}}
\setcounter{footnote}{0}

\section{Introduction}

Conformal invariance is an important device to probe diverse area of theoretical physics and its application to gravity theory was initiated
with the idea that
 conformally invariant gravity in four dimensions  \cite{Weyl:1918ib}
 might result in an unified description of gravity and electromagnetism, thus extending the Einstein's general relativity.
 It is well-known that the Einstein-Hilbert action of general relativity is not a conformally invariant theory and in order to
realize the conformal invariance, one has to resort to curvature-squared  gravity with Weyl tensor, or to introduce conformal scalar fields  \cite{Dirac:1938mt} which compensate the conformal
 transformation of the metric. In the latter Weyl gravity  in which we are interested, a quartic potential for the scalar field
 can be allowed and its higher dimensional extensions are straightforward \cite{Moon:2009zq}.

In yet alternative attempt to apply the conformal symmetry to gravity,
an anisotropic extension in which the homogeneous conformal transformation
is generalized to anisotropic scaling symmetry can be considered. The prototype gravity theory which exhibits such a feature is the  Horava-Lifshitz (HL) gravity \cite{Horava:2009uw} which was proposed as a quantum theory of gravity.
 Essential feature of HL gravity is the anisotropic scaling of
space and time as a fundamental symmetry of quantum gravity, thus
abandoning the Lorentz symmetry at short distance.
This idea was extended \cite{Papantonopoulos:2011xa} to the five dimensional theory in which the higher dimensional theory breaks diffeomorphism invariance explicitly at the action level  to its foliation-preserving subgroup. The foliation can be adapted to an extra space dimension, thus leaving the four-dimensional spacetime diffeomorphism intact\footnote{The action considered in the paper \cite{Papantonopoulos:2011xa} (Eq. (2.3) of the reference), is not the most general action with foliation preserving diffeomorphism
symmetry containing up to quadratic derivatives. Compare with our Eq. (\ref{conformalR}).
}.

The purpose of this work is to construct $4+D$ dimensional  conformal gravity in which the action is invariant under anisotropic conformal transformations:
the spacetime (4) and extra dimensional ($D$) metric enjoy conformal transformations with different conformal weights. This kind of anisotropic conformal transformation was considered before in an attempt to extend the original  Weyl invariance of Horava-Lifshitz gravity by using an extra scalar field.
An anisotropic Weyl invariant theory whose action is invariant
under the anisotropic conformal transformations of the space and time metric components in the ADM formalism was constructed and its cosmological consequences were investigated \cite{Moon:2010wq}.

In extension to higher dimensional space-time and construction of the anisotropic invariant gravity, there are essentially two steps that can be taken. Since we are treating the spacetime and extra dimensions on a different footing, we do not require the full diffeomorphism
symmetry on $4+D$ dimensions  and  in order to implement the anisotropy between them, we only demand  the foliation preserving diffeomorphism (FPD) invariance where the foliation is adapted along the extra dimensions so that the spacetime
4D general covariance is intact.
The next step is to contemplate the conformal transformations of the spacetime and extra dimensional metrics with  a scalar field that compensates the conformal weight of the metrics. Here, a parameter which describes the degree of anisotropy of conformal transformation between the spacetime and extra dimensional metrics can be introduced.  It will be denoted by $z$ which becomes a dynamical  exponent of the action.  After these steps, one arrives at a conformally anisotropic higher dimensional gravity  but which preserves spacetime diffeomorphism.
Since the action breaks $4+D$ general covariance explicitly, there may arise instability due to ghost-like excitations in the quadratic perturbation around a vacuum. This is checked in detail in 5D case. Another point is that a priori  an arbitrary real value of $z$
can be assigned for the endowment  of anisotropic conformal invariance and the resulting theory encompass various limit including Einstein and conformal Weyl gravity,
but some specific value of $z$ could be preferred.

The paper is organized as follows:
In Sec. 2, we explicitly construct $4+D$ dimensional gravity with  FPD  invariance. In Sec. 3, it is extended to   anisotropic conformal invariance
and discussions on the dimensional reduction to effective four dimensional gravity are given.  In Sec. 4, a more detailed account is given for    5D case.  Some exact solutions are given and  perturbations of the quadratic action are presented.  Sec. 5 includes  summary and  possible applications to the cosmological constant problem with a specific choice of $z$.

\section{FPD invariant gravity}

Let us consider $(d+D)$ dimensional spacetime where the foliation is adapted to $D$ dimensions. The $(d+D)$ dimensional metric can be written as
\begin{eqnarray}\label{mett}
ds^2=g_{AB}(x,y)(dx^{A} +N^{A}_{M} dy^M)(dx^B +N^{B}_{N}
dy^N)+\gamma_{MN}(x,y)dy^M dy^N, \label{GADM}
\end{eqnarray} where  $A,~B,\cdots$ indices
 denote those for $d$-dimensional base
spacetime with coordinate $x$ and $M,~N$  are those for $D$-dimension
with coordinate $y$. Here, $N_M^{A}$ is a shift vector along the base spacetime corresponding to the $M$-direction \cite{Misner:1974qy}, $g_{AB}$ and $\gamma_{MN}$ correspond to $d$ and
$D$-dimensional metric, respectively.

The above equation (\ref{GADM}) is a generalized ADM decomposition:
 (i) When ${\rm A,B},\cdots=1,2,3$ and  ${\rm M ,N,}\cdots=0$,
 it is the usual ADM decomposition. (ii)  ${\rm A,B},\cdots=4,5,\dots d+3$ and  ${\rm M ,N,}\cdots=0,1,2,3$; Extra dimensions are foliated along the spacetime and it is the usual Kaluza-Klein reduction \cite{Appelquist:1987nr}.
 (iii) The case we are interested in is when $4$ dimensional hypersurface is foliated along  the extra $D$-dimensions, which is given by ${\rm A,B},\cdots=0,1,2,3$ and  ${\rm M ,N,}\cdots=4,5,\dots D+3$. From here on we consider $d=4.$ (Also, change the notation ${\rm A,B},\cdots \rightarrow \mu,\nu,\cdots$ and  ${\rm M ,N,}\cdots\rightarrow m,n,\cdots$.)
To proceed, we consider foliation preserving diffeomorphism:
\begin{eqnarray}\label{xytr}
x^{\mu}\to x^{\prime\mu}\equiv x^{\prime\mu}(x,y),~~~y^n\to y^{\prime n}\equiv
y^{\prime n}(y),
\end{eqnarray}
whose infinitesimal transformations are given by
\begin{eqnarray}
x^{\prime \mu}=x^\mu+\xi^\mu(x,y),~~~y^{\prime m}=y^{m}+\eta^m(y);~~~\partial_{\mu}
\eta^m=0.
\end{eqnarray}
Here $\eta^m(y)$ is a function of $y^m$ only. One can  check that
\begin{eqnarray}
\delta g_{\mu\nu}&=&-{\cal L}_{\xi}(x)g_{\mu\nu}-\eta^m(y)\partial_{m}g_{\mu\nu}
\label{inft1},\\
\delta N^{\mu}_{m}&=&-{\cal L}_{\xi}(x)N^{\mu}_{m}-
{\cal L}_{\eta}(y)N^{\mu}_{m}-\partial_m\xi^{\mu},\label{inft2}\\
\delta \gamma_{mn}&=&-{\cal
L}_{\eta}(y)\gamma_{mn}-\xi^\mu\partial_{\mu}\gamma_{mn},\label{inft3}
\end{eqnarray}
where ${\cal L}_{\xi}(x)$ and ${\cal L}_{\eta}(y)$ with
$\xi=\xi^{\mu}\partial_{\mu},~\eta=\eta^{m}\partial_{m}$ are the
Lie-derivative acting only on the indices of $x$ and $y$,
respectively.

From the transformations (\ref{inft1})-(\ref{inft3}), one obtains finite coordinate
transformations as follows
\begin{eqnarray}
g^{'}_{\mu\nu}(x',y')&=&\frac{\partial x^\rho}{\partial
x^{'\mu}}\frac{\partial
x^\sigma}{\partial x^{'\nu}}g_{\rho\sigma}(x,y),\label{trans1}\\
{N'}_{m}^{\mu}(x',y')&=&\Big(\frac{\partial y^n}{\partial
y^{'m}}\Big)\Big[\frac{\partial x^{'\mu}}{\partial
x^{\nu}}N_{n}^{\nu}(x,y)-\frac{\partial x'^{\mu}}{\partial
y^{n}}\Big],\label{trans2}\\
\gamma'_{mn}(x',y')&=&\frac{\partial y^p}{\partial
y^{'m}}\frac{\partial y^q}{\partial y^{'n}}\gamma_{pq}(x,y).\label{trans3}
\end{eqnarray}
One can check that the above transformations (\ref{trans1})-(\ref{trans3}) leave the line element (\ref{GADM}) invariant.
The next step is to construct Diff(4)-covariant derivative and field strength \cite{Cho:1991wn, Yoon:2000sq} \footnote{ FPD in Ref. \cite{Cho:1991wn, Yoon:2000sq} is restricted to the case where $y^{\prime n}=
y^{n}, \eta^n=0$}:
\begin{eqnarray}\label{ddef}
{\cal D}_{m}g_{\mu\nu}&=&\partial_m g_{\mu\nu}-{\cal
L}_{\xi_m}(x)g_{\mu\nu}~~( ~ \xi_m=N^\mu_m \frac{\partial}{\partial x^\mu})\nn\\
&=&\partial_{m}g_{\mu\nu}-N^{\rho}_{m}\partial_{\rho}g_{\mu\nu}-
(\partial_{\mu}N^{\rho}_{m})g_{\rho\nu}
-(\partial_{\nu}N^{\rho}_{m})g_{\mu\rho},\\
{\cal F}_{mn}&=&[{\cal D}_m,{\cal D}_{n}],~~(\equiv {\cal F}^{\mu}_{mn}\partial_\mu) \nn\\
{\cal F}^{\mu}_{mn} &=&\partial_{m}N_{n}^{\mu}-\partial_{n}N_{m}^{\mu}-N_{m}^{\nu}\partial_{\nu}N_{n}^{\mu}
+N_{n}^{\nu}\partial_{\nu}N_{m}^{\mu}.\label{fdef}
\end{eqnarray}
One can  check that under FPD (\ref{xytr}), the above
quantities (\ref{ddef}), (\ref{fdef}) transform covariantly  as
\begin{eqnarray}\label{dftr}
({\cal D}_m g_{\mu\nu})^{'}&=&\frac{\partial y^{n}}{\partial
y^{'m}}\frac{\partial x^{\rho}}{\partial x^{'\mu}}\frac{\partial
x^{\sigma}}{\partial x^{'\nu}}{\cal D}_n g_{\rho\sigma},\label{inva1}\\
({\cal F}^{\mu}_{mn})^{'}&=&\frac{\partial y^{p}}{\partial
y^{'m}}\frac{\partial y^{q}}{\partial y^{'n}}\frac{\partial
x^{'\mu}}{\partial x^{\nu}}{\cal F}^{\nu}_{pq}.\label{inva2}
\end{eqnarray}
One also has the derivative of the metric $\gamma_{mn}$ transforming as
\begin{eqnarray}
(\partial_\mu \gamma_{mn})^{'}&=&\frac{\partial x^{\nu}}{\partial
x^{'\mu}}\frac{\partial y^{p}}{\partial y^{'m}}\frac{\partial
y^{q}}{\partial y^{'n}}(\partial_\nu \gamma_{pq}).\label{inva3}
\end{eqnarray}
under FPD (\ref{xytr}). The covariant quantities (\ref{inva1})-(\ref{inva2})
can be contracted to generate FPD invariant  quantities.
Geometrically,  ${\cal D}_m g_{\mu\nu}$ is a generalized extrinsic curvature along $m$-th extra dimension and ${\cal F}^{\mu}_{mn}$ is a Diff(4) vector-valued curvature associated with the holonomy
of the shift vector $N_n^\mu$ along the extra dimensions.

Other invariants are  curvatures of $4$ dimensional spacetime and $D$ extra dimensions. In order to construct these, let us first define
\begin{equation}
\hat\partial_m= \partial_m -N_m^\mu\partial_\mu.
\end{equation}
Then, we note that under FPD (\ref{xytr}),
\begin{equation}
\hat{\partial}^{'}_m=\left(\frac{\partial y^n}{\partial y^{'m}}\right)
\hat\partial_n,~~~~~
{\partial}^{'}_{\mu}=\left(\frac{\partial x^{\nu}}{\partial x^{'\mu}}\right)
\partial_{\nu}. ~
\end{equation}
Therefore, from (\ref{trans1}),  the curvature
scalar  $R^{(4)}=g^{\mu\nu}R_{\mu\nu}$ is a FPD invariant. Likewise,
$\hat{R}^{(D)}=\gamma^{mn}\hat{R}_{mn}$, where
 $\hat{R}_{mn}$ is defined as
\begin{eqnarray}
\hat{R}_{mn}=\partial_{p}\hat{\Gamma}_{nm}^{p}-\partial_{n}\hat{\Gamma}_{pm}^{p}+\hat{\Gamma}_{pq}^{q}\hat{\Gamma}_{nm}^{p}
-\hat{\Gamma}_{pn}^{q}\hat{\Gamma}_{qm}^{p},
\end{eqnarray}
with the connection $\hat{\Gamma}_{mn}^{p}$ given by
\begin{eqnarray}
\hat{\Gamma}_{mn}^{p}=\frac{1}{2}\gamma^{pq}(\hat{\partial}_m\gamma_{nq}
+\hat{\partial}_n\gamma_{qm}-\hat{\partial}_q\gamma_{mn})
\end{eqnarray}
is also a FPD  invariant.
Consequently, we can construct a general FPD invariant $(4+D)$ dimensional action as follows:
\begin{eqnarray}\label{conacd}
S_{(\rm{FPD})}&=&\int d^{4+D}x \sqrt{-g}M_*^{2+D}\sqrt{\gamma}\Big[
\left( R^{(4)}-2\Lambda\right)+\alpha_{1}\hat{R}^{(D)}-\frac{\alpha_2}{4}\gamma^{mn}\gamma^{pq}g_{\mu\nu}F_{mp}^{\mu}F_{nq}^{\nu}
\nn\\
&& -\frac{\alpha_3}{4}\gamma^{mn}g^{\mu\nu}g^{\alpha\beta}({\cal
D}_m g_{\mu\alpha}{\cal D}_{n}g_{\nu\beta}
 -\alpha_4{\cal D}_m g_{\mu\nu}{\cal D}_{n}g_{\alpha\beta})\nonumber\\
 &&-\frac{\alpha_5}{4}g^{\mu\nu}\gamma^{mn}\gamma^{pq}(\partial_{\mu}
\gamma_{mp}\partial_{\nu}\gamma_{nq}
 -\alpha_6\partial_{\mu}
\gamma_{mn}\partial_{\nu}\gamma_{pq})\Big] +S_{m},\label{ehq}
\end{eqnarray}
where $\alpha_{1\sim6}$ are arbitrary constants and $M_*$ is the higher dimensional gravitational constant.
One can check that when $\alpha_{1\sim6}=1$, the action (\ref{conacd})
combines into  the $(4+D)$ dimensional Einstein-Hilbert action with cosmological constant with the FPD being elevated to the $4+D$ general covariance.
 $S_m$ is the matter action whose constituents will not be our concern, but we will compute general constraint on the energy-momentum tensor of $S_m$ imposed by FPD invariance in 5D case.
 For generic values of $\alpha$'s  the theory exhibits pathological behaviors displaying ghost-like excitations in the perturbation theory \cite{Papantonopoulos:2011xa}. This issue will be taken up later  for the  5D conformally invariant case. 

\section{Conformal extension}

In this section, we first consider anisotropic extension of the action (\ref{conacd}), whose action is invariant
with respect to the anisotropic conformal transformations:
\begin{eqnarray}\label{weyltf}
g_{\mu\nu}\rightarrow e^{2\omega(x,y)}g_{\mu\nu},~~
\gamma_{mn}\rightarrow e^{2z\omega(x,y)}\gamma_{mn},~~ \phi\rightarrow
e^{-v(z)\omega(x,y)}\phi,~~N_{m}^{\mu}\rightarrow N_{m}^{\mu}.\label{confotrans1}
\end{eqnarray}
Here, we have introduced a parameter $z$ which characterizes the anisotropy of the extra dimension and $v(z)$ is an arbitrary function of $z$. $z=1$ and $v(z)=1+D/2$ corresponds to the isotropic conformal transformations
\cite{Moon:2009zq}.
Taking into account an additional scalar field $\phi$, one can construct the anisotropic invariant action in ($4+D$) dimensions as
\begin{eqnarray}
&&\hspace*{-1em}S_{(C)}=\int d^{4+D}x \sqrt{-g^{(4)}}\sqrt{\gamma}M_*^{2+D}\Bigg[
 \phi^{\frac{2+Dz}{v}}\Big(R^{(4)}+A_1\frac{\nabla_{\mu}\nabla^{\mu}\phi}{\phi}
 +A_2\frac{\nabla_{\mu}\phi\nabla^{\mu}\phi}{\phi^2}\Big)-V_0\phi^{2N}\nn\\
 &&+\alpha_1\phi^{\frac{4+(D-2)z}{v}}\Big(\hat{R}^{(D)}+
B_1\frac{\hat{\nabla}_{m}\hat{\nabla}^{m}\phi}{\phi}+B_2\frac{\hat{\nabla}_{m}\phi\hat{\nabla}^{m}\phi}{\phi^2}\Big)
-\frac{\alpha_2}{4}\phi^{\frac{6+(D-4)z}{v}}\gamma^{mn}\gamma^{pq}g_{\mu\nu}F_{mp}^{\mu}F_{nq}^{\nu}\nn\\
 &&-\alpha_3\phi^{\frac{4+(D-2)z}{v}}\Bigg\{\frac{1}{4}\gamma^{mn}g^{\mu\nu}g^{\alpha\beta}({\cal
D}_m g_{\mu\alpha}{\cal D}_{n}g_{\nu\beta}
 -\alpha_4{\cal D}_m g_{\mu\nu}{\cal D}_{n}g_{\alpha\beta})\nonumber\\
&&\hspace*{15em}+C_1\gamma^{mn}g^{\mu\nu}{\cal D}_m
g_{\mu\nu}\frac{\hat{\nabla}_n\phi}{\phi}
+C_2\gamma^{mn}\frac{\hat{\nabla}_m\phi\hat{\nabla}_n\phi}{\phi^2}\Bigg\}\nn\\
&&-\alpha_5\phi^{\frac{2+Dz}{v}}\Bigg\{\frac{1}{4}g^{\mu\nu}\gamma^{mn}\gamma^{pq}(\nabla_{\mu}
\gamma_{mp}\nabla_{\nu}\gamma_{nq}
 -\alpha_6\nabla_{\mu}
\gamma_{mn}\nabla_{\nu}\gamma_{pq})\nonumber\\
&&\hspace*{15em}+D_1g^{\mu\nu}\gamma^{mn}\nabla_{\mu}\gamma_{mn}\frac{\nabla_{\nu}\phi}{\phi}
+D_2g^{\mu\nu}\frac{\nabla_{\mu}\phi\nabla_{\nu}\phi}{\phi^2}\Bigg\}\Bigg],
\label{confact1}
\end{eqnarray}
where $M_*$ sets up a higher dimensional gravity scale and the Weyl scalar $\phi$ is dimensionless
field \footnote{Note that the over-all scale $M_*$  cannot be absorbed into the scalar field all together in (\ref{confact1}), unless $z=1$ so that $v(z)=1+D/2$, i.e., the isotropic case. From here on, we
 choose $v=(2+Dz)/2$, because it simply yields the conventional 4D
non-minimal coupling term of $\phi^2 R^{(4)}$. The case $Dz=-2$ is given a separate treatment.}. The hat covariant derivative is given by
$\hat{\nabla}_m j^m=\hat{\partial}_m j^m+\hat{\Gamma}_{mp}^m
j^p$ and the coefficients $A_{1,2},~B_{1,2},~C_{1,2},$ $D_{1,2}$ are
fixed as
\begin{eqnarray}\label{abcd}
&&\hspace*{-2em}A_1=-\frac{6}{v},~~~~A_2=-\frac{6}{v}\Big(\frac{1}{v}-1\Big),~~~~B_1=\frac{2z(1-D)}{v},~~~~B_2=\frac{2z(1-D)}{v}
\Big(\frac{z(D-2)}{2v}-1\Big),\nonumber\\
&&\hspace*{-2em}C_1=-\frac{1-4\alpha_4}{v},~~~~C_2=-\frac{4(1-4\alpha_4)}{v^2},
~~~~D_1=\frac{z(1-\alpha_6 D)}{v},~~~~D_2=\frac{z^2D(1-\alpha_6
D)}{v^2}.
\end{eqnarray}
Also, the exponent $N$ of the scalar field can be found as
\begin{eqnarray}
N=\frac{4+Dz}{2v}.
\end{eqnarray}
When $\alpha_{1\sim6}=1$ and $z=1$, the above action (\ref{confact1})
combines into the $(4+D)$ dimensional conformally invariant gravity with $4+D$ general covariance \cite{Moon:2009zq};
\begin{eqnarray}
S_{(CW)} = \int  \,d^nx  \sqrt{-g^{(4+D)}} \left\{\phi^2
R^{(n)}+\frac{4(n-1)}{n-2}g^{ab}\nabla_{a}\phi\nabla_{b}\phi-V_{0}
\phi^{\frac{2n}{n-2}}\right\},\label{s3}
\end{eqnarray}
where $a,b=0, \cdots n-1$ and $n=4+D.$

In passing, we make the following remarks. The only dimensionful
parameter inside the parenthesis of (\ref{confact1}) is $V_0$ which
has mass squared dimensions originating from the cosmological
constant term in (\ref{ehq}). Also, the action (\ref{confact1}) has
conformal invariance for any real values of $z. $ That negative
values of $z$ might lead to some interesting application regarding
the cosmological constant will be discussed later. Finally, it is to
be noticed that FPD (\ref{trans1})-(\ref{trans3})  remains as
symmetry along with the scalar transformation $\phi^\prime(x^\prime,
y^\prime)=\phi(x,y)$ in the action (\ref{confact1}), and FPD
invariance and conformal symmetry are compatible with each other.

Now we discuss 4-dimensional effective action and let us consider only
zero modes \cite{Appelquist:1987nr}.
It is convenient to go to a frame where $N_m^{\mu}=0$ (``comoving'' frame) and we impose $y$-independence (cylindrical condition) for $g_{\mu\nu}=g_{\mu\nu}(x), \phi=\phi(x)$ and $\gamma_{mn}=\gamma_{mn}(x)$.
In this approximation, the action (\ref{confact1}) can be expressed as
\begin{eqnarray}\label{actn}
&&\hspace*{-1em}S_{(C)}=\frac{1}{16\pi G'}\int d^{4}x \sqrt{-g^{(4)}}\vert\rm{det}{\gamma}(x)\vert^{\frac{1}{2}}\Bigg\{
 \phi^{2} R^{(4)}
 +\bar Ag^{\mu\nu}{\nabla_{\mu}\phi\nabla_{\nu}\phi}-V(\phi)\nn\\
&&\hspace*{-1em}-\alpha_5\phi^{2}\Big[\frac{1}{4}g^{\mu\nu}\gamma^{mn}\gamma^{pq}(\nabla_{\mu}
\gamma_{mp}\nabla_{\nu}\gamma_{nq}
 -\alpha_6\nabla_{\mu}
\gamma_{mn}\nabla_{\nu}\gamma_{pq})\Big]+\bar D \gamma^{mn}(\nabla_{\mu}\gamma_{mn})\phi{\nabla^{\mu}\phi}
\Bigg\}.
\label{confact2}
\end{eqnarray}
The constant $G'$ and the potential are given  by
\begin{equation}
\frac{1}{16\pi G'}=M_*^{2+D}V(D), ~~V(\phi)=V_0\phi^{2+\frac{4}{2+Dz}},\label{redpot1}
\end{equation}
where $V(D)$ is the volume of the extra dimensions, $V(D)=\int d^Dy.$ And

\begin{equation}
\bar A=\frac{1}{v^2}\left[ 12v-6-\alpha_5z^2D(1-\alpha_6D)\right],~~
 \bar D=\frac{1}{v}\left[3-\alpha_5z(1-\alpha_6D)\right].\label{onesix}
\end{equation}
 There are several interesting aspects worthy of discussions in the above action (\ref{confact2}).  The power of the potential is exotic and can become any, in principle. $z=0$ gives the usual $\phi^4$ potential.
To avoid possible discontinuities of the potential and to have the parity symmetry $\phi\rightarrow -\phi$, we can restrict to only even powers of potential which selects $z$ as
\begin{equation}
z=\frac{2}{D}\left(\frac{1}{N-1}-1\right)~~   (N= \rm{integers}),\label{parity}
\end{equation}
with potential of the form $\sim \phi^{2N}$ with positive as well as negative powers being allowed.
The $\gamma_{mn}$ fields  behave like 4D scalar fields under general covariance (i.e., Eq. (\ref{trans3}))
 in the above action and it  describes scalar-tensor theory of gravity coupled with  a complex nonlinear sigma model type given by  $\gamma_{mn}$'s. It will be demonstrated explicitly that the gravity sector of the action (\ref{confact2})
has the Einstein-Hilbert action at one side and Weyl gravity at the other. The existence of  ghost-like excitations of $\gamma$'s is dependent upon the  generic values of $\alpha$'s.
For example, when $\alpha_5=\alpha_6=1$, this problem can be avoided (see below).

 It is crucial to observe  that  the action (\ref{confact2})
exhibits residual 4D conformal invariance
\begin{equation}
g_{\mu\nu}\rightarrow e^{2\omega(x)}g_{\mu\nu},~~
\gamma_{mn}\rightarrow e^{2z\omega(x)}\gamma_{mn},~~ \phi\rightarrow
e^{-(1+\frac{Dz}{2})\omega(x)}\phi,
\end{equation}
where $\omega$ is now a function of $x$ only, $\omega\equiv \omega(x).$
The residual conformal invariance enables to fix further the conformal scalar to be $\phi=\phi_0$, and the resultant action can be expressed as
\begin{eqnarray}
\hspace*{-1em}S_{(\gamma)}&=&\frac{1}{16\pi G_N}\int d^{4}x \sqrt{-g}\vert\rm{det}{\gamma}(x)\vert
^{\frac{1}{2}}\Bigg\{
 \Big(R^{(4)}-2\Lambda_\gamma\Big)\nonumber\\
&&~~~~~~~~~~~~~~~~~~~~-\alpha_5\Big[\frac{1}{4}g^{\mu\nu}\gamma^{mn}\gamma^{pq}(\nabla_{\mu}
\gamma_{mp}\nabla_{\nu}\gamma_{nq}
 -\alpha_6\nabla_{\mu}
\gamma_{mn}\nabla_{\nu}\gamma_{pq})\Big]\Bigg\},\label{chofreund}
\end{eqnarray}
where
\begin{equation}
\frac{1}{16\pi G_N}=M_*^{2+D}\phi_0^2V(D), ~~\Lambda_\gamma=\frac{1}{2}V_0\phi_0^{\frac{4}{2+Dz}}.\label{cc1}
\end{equation}
We remark that we would have arrived at the same action, had we taken the alternative route: We can first fix $\phi=\phi_0$ using the conformal invariance in the action (\ref{confact1}).  Then, restrictions  to zero modes again with only $x$-dependence lead to an effective 4D theory which preserves 4D general covariance. In the comoving frame, the action assumes the same form as (\ref{chofreund}). Note that the residual symmetry is independent of the cylindrical condition: After fixing $\phi=\phi_0,$ the action (\ref{confact1}) still respects $4+D$ FPD (\ref{trans1})-(\ref{trans3}). Alternatively,  in the comoving gauge with $N_m^\mu=0$,  the action  (\ref{confact1}) is still invariant under the conformal transformation (\ref{confotrans1}).
When $\alpha_5=\alpha_6=1$, the action  (\ref{chofreund}) coincides with the zero mode sector of the scalar-graviton action in Ref. \cite{Cho:1975sf} with an extra  cosmological constant given by Eq. (\ref{cc1}), and in this case, there is no instability problem.
Finally, we remark that one  can check that   the above action (\ref{chofreund}) admits solution given by
\begin{equation}
R^{(4)}_{\mu\nu}=\Lambda_{\gamma} g_{\mu\nu},~~\gamma_{mn}=\gamma^0_{mn}(=\rm{const}).
\end{equation}
This solution describes 4D de Sitter space when $\Lambda_\gamma>0.$\footnote{The solution may not be the exact solutions of full $(4+D)$-dimensional field equations. However, this point will not be pursued further in this work.}

We can go one step further to absorb the factor $\vert\rm{det}{\gamma}(x)\vert
^{\frac{1}{2}}$ in (\ref{chofreund}) into the metric by  using the conformal transformation  via
\begin{equation}
\tilde g_{\mu\nu}=\vert{\rm det}{\gamma}(x)\vert
^{\frac{1}{2}}g_{\mu\nu}.
\end{equation}
Rewriting (\ref{chofreund}) in terms of $\tilde g_{\mu\nu}$, we obtain
(with $\tilde g_{\mu\nu}\rightarrow g_{\mu\nu})$
\begin{eqnarray}
S^\prime_{(\gamma)}&=&\frac{1}{16\pi G_N}\int d^{4}x
\Bigg\{
 R^{(4)}-2\vert{\rm det}{\gamma}(x)\vert
^{-\frac{1}{2}}\Lambda_\gamma\nonumber\\
&&\hspace*{7em}
-\alpha_5\Big[\frac{1}{4}g^{\mu\nu}\gamma^{mn}\gamma^{pq}(\nabla_{\mu}
\gamma_{mp}\nabla_{\nu}\gamma_{nq}
 -\alpha_6^\prime\nabla_{\mu}
\gamma_{mn}\nabla_{\nu}\gamma_{pq})\Big]\Bigg\},\nn
\end{eqnarray}
with $\alpha_6^\prime=\alpha_6-\frac{3}{2\alpha_5}.$
The above action describes the  Einstein-Hilbert action coupled to nonlinear sigma model with target space coordinates given by $\gamma_{mn}$'s with a potential term. In 5D, the nonlinear sigma model is simply a scalar field theory with an exponential potential ($\Phi\equiv \log \gamma_{55}$);
\begin{equation}
V(\Phi)=2\Lambda_\gamma \exp(-\frac{1}{2}\Phi).
\end{equation}
The scalar field is canonical or ghost depending on $\alpha_5+\frac{3}{2}-\alpha_5\alpha_6>0$ or $<0$ as can be readily checked.

There exists other route as far as gauge fixing is concerned. Instead of fixing $\phi$, we can fix one component of $\gamma_{mn}$ or a combination of $\gamma_{mn}$'s to be a constant. The most natural choice from the action (\ref{confact2}) is put $\vert\rm{det}{\gamma}(x)\vert=1.$ Then, we end up with
\begin{eqnarray}\label{actn}
&&\hspace*{-1em}S_{(\phi, \gamma)}=\frac{1}{16\pi G'}\int d^{4}x \sqrt{-g}\Bigg\{
 \phi^{2} R^{(4)}
 +\bar Ag^{\mu\nu}{\nabla_{\mu}\phi\nabla_{\nu}\phi}-V(\phi)
 \nn\\
&&~~~~~~~~~~~~~~~~~~~-\alpha_5\phi^{2}\Big[\frac{1}{4}g^{\mu\nu}\gamma^{mn}\gamma^{pq}\nabla_{\mu}
\gamma_{mp}\nabla_{\nu}\gamma_{nq}
\Big]+\lambda(\vert\rm{det}{\gamma}(x)\vert-1)
\Bigg\},\label{chofreund2}
\end{eqnarray}
where $\lambda$ is a Lagrange multiplier. Now, the gravity sector is given by the 4D Weyl gravity\footnote{In five dimensional case with $\gamma_{55}=N^2$, $\alpha_5$ term vanishes and only the gravity sector survives.}. There are, however,    noticeable changes from the ordinary 4D   Weyl gravity. The coefficient $\bar A$ is no longer the well-known ${6}$, but is given by the value in (\ref{onesix}).  Recall that in 4D, conformal invariance would inevitably render $\phi$ to be a ghost with a positive value of 6.  In our case, there arise additional contributions from the conformal invariance of $\gamma_{mn}$ sector which could add up to make the overall coefficient $\bar A$ to be negative, making $\phi$ carry positive kinetic energy for certain range of parameters. This renders the resulting 4D action to be the Brans-Dicke gravity.  Another difference is that the power of the potential is drastically different from   4.  These properties are basically inherited from anisotropic conformal invariance of the higher dimensions. It is interesting to observe that both the Einstein-Hilbert action and Brans-Dicke gravity can originate as two different gauged fixed versions from the conformally invariant action (\ref{confact2}).

 A couple of comments are in order. Conformal invariance alone does not put any restrictions on  the value of $z$. When $Dz=-2,$ one can not choose $v= 1+ \frac{Dz}{2}$ and this case has to be treated separately. Nevertheless, the anisotropically conformal invariant action is still given by Eq. (\ref{confact1}). If we impose $Dz=-2 $ in the action (\ref{confact1}), it assumes a similar form as (\ref{confact1}) with the noticeable difference that the conformal factor $\phi^2$ is no longer multiplying $R^{(4)}$. If we perform the dimensional reduction by imposing the cylindrical condition and fixing $\phi=\phi_0$ as before, we end up with the exactly same forms as (\ref{chofreund})   with ($v=1$)
\begin{equation}
\frac{1}{16\pi G_N}=M_*^{2+D}V(D), ~~\Lambda_{(2)}=\frac{1}{2}V_0\phi_0^{2}.\label{cc2}
\end{equation}
Likewise, fixing  $\vert\rm{det}{\gamma}(x)\vert=1$ yields action of the type (\ref{chofreund2})
in which the Weyl $\phi^2 R^{(4)}$ term is replaced by the Einstein-Hilbert $R^{(4)}$ and the potential is quadratic, $V\sim \phi^2$. It can be readily checked that  this $Dz= -2$ case corresponds to redefinition of $N\rightarrow\infty$  limit  of the action (\ref{confact1})   with the parity symmetry $\phi\rightarrow-\phi$ being imposed as in (\ref{parity}).
Another comment is that the action (\ref{confact1}) does not fit to describe 5D case, since scalar curvature $R^{(D=1)}$ and $F_{mn}^\mu$ are always zero. Therefore, separate investigations of anisotropic invariant action in 5D are performed in the next section, in which we explore other  ground states  relaxing the cylindrical condition and considering ground states with $y$-dependence.  We  also perform the quadratic perturbations to check the ghost-like excitations.


\section{$z$-Weyl gravity in 5D}

In this section, we investigate 5D case in some detail and include $S_m$ later for generality. We search solutions of equations of motion and perform quadratic perturbations around a vacuum to check the instability.
We intend to make this section be self-contained. For that purpose, let us first rewrite  the anisotropic invariant action in 5D, whose form with slight adjustments of notations  and rearrangements of terms
(see below) can be expressed as
\begin{eqnarray}
&&\hspace*{-2em}S_{(5{\rm D})}=\int dy
d^4xN\sqrt{g_{(4)}}~M_{*}^{3}\Bigg[\varphi^{2}\left(R^{(4)}
-\frac{12}{z+2}\frac{\nabla_{\mu}\nabla^{\mu}\varphi}{\varphi}
+\frac{12z}{(z+2)^2}\frac{\nabla_{\mu}\varphi\nabla^{\mu}\varphi}{\varphi^2}
\right)\nonumber\\
&&\hspace*{7em}+
\beta_1\varphi^{-\frac{2(z-4)}{z+2}}\left\{B_{\mu\nu}B^{\mu\nu} -\lambda
B^2\right\}+\beta_2 \varphi^{2}A_{\mu}A^{\mu}-V(\varphi)\Bigg],\label{conformalR}
\end{eqnarray}
where the potential $V$ is
\begin{eqnarray}\label{pote}
V=V_0\varphi^{\frac{2(z+4)}{z+2}}.\label{fracpote}
\end{eqnarray}
In the action (\ref{conformalR}), $B_{\mu\nu}$ and $A_{\mu}$ are given by
\begin{eqnarray}
B_{\mu\nu}&=&K_{\mu\nu}+\frac{2}{(z+2)N\varphi}g_{\mu\nu}(\partial_y
\varphi-\na_{\rho}\varphi N^{\rho}),\\
A_{\mu}&=&\frac{\partial_{\mu}N}{N}
+\frac{2z}{z+2}\frac{\partial_{\mu}\varphi}{\varphi},
\end{eqnarray}
where $K_{\mu\nu}$ is the extrinsic curvature tensor,
$K_{\mu\nu}=(\partial_y g_{\mu\nu} - \nabla_{\mu} N_{\nu} -
\nabla_{\nu} N_{\mu})/(2N)$.
Note that the above action (\ref{conformalR}) can be obtained from the action (\ref{confact1}), when replacing $\phi$ with $\varphi$ and choosing $D=1$, $v=(2+z)/2$, $\alpha_3=-\beta_1$, $\alpha_4=\lambda$, $\alpha_5(\alpha_6-1)=\beta_2$.
 When $\beta_1=\lambda=z=1$ and $\beta_2=0$  it is 5D Weyl gravity, Eq. (\ref{s3}) with $D=1$.  Note also that the action (\ref{conformalR}) is simply the anisotropic invariant extension of the following action considered in Ref. \cite{Papantonopoulos:2011xa}\footnote{$\beta_2$ term is omitted in this reference.}:
\begin{eqnarray}
S_{(\rm{}5F)}=\int \,dy
d^4xN\sqrt{g_{(4)}}~\left[\left(R^{(4)}-2\Lambda_5\right)+\beta_1\{K_{\mu\nu}K^{\mu\nu} -\lambda
K^2\}+\beta_2N^{-2}g^{\mu\nu}\partial_{\mu}N\partial_{\nu}N\right],
\nonumber\label{5dR}
\end{eqnarray}
obtained from the action (\ref{conacd}) in 5D. One can check that
the action (\ref{conformalR}) is anisotropically conformal invariant
with respect to \bea N&\rightarrow&
e^{z\omega}N,~~~N_{\mu}\rightarrow e^{2\omega}N_{\mu},\nn\\
g_{\mu\nu}&\rightarrow& e^{2\omega}g_{\mu\nu},~~\varphi\rightarrow
e^{-\frac{z+2}{2}\omega}\varphi, \label{trans}\eea
where $\omega=\omega(x,y)$.

Now we compute the equations of motion for the action (\ref{conformalR}).
Varying for $N$, $N^{\nu}$, $g^{\mu\nu}$, $\varphi$ in the action
(\ref{conformalR}) leads to the equations of motion:
\begin{eqnarray}
&&\hspace*{-3em}\delta_{N}S_{5{\rm D}};~~\varphi^{2}\left(R_{(4)}
-\frac{12}{z+2}\frac{\nabla_{\mu}\nabla^{\mu}\varphi}{\varphi}
+\frac{12z}{(z+2)^2}\frac{\nabla_{\mu}\varphi\nabla^{\mu}\varphi}{\varphi}
\right)+\beta_1\varphi^{-\frac{2(z-4)}{z+2}}\left(B_{\mu\nu}B^{\mu\nu}
-\lambda B^2\right)\nonumber\\
&&\hspace*{3em}-\beta_2\varphi^{2}(A_{\mu}A^{\mu}+2\nabla_{\mu}A^{\mu}
+\frac{8}{z+2}\frac{\nabla_{\mu}\varphi }{\varphi}A^{\mu})-V(\varphi)=0,~~~\label{Neq}\\
&&\hspace*{-3em}\delta_{N^{\nu}}S_{5{\rm D}};~~\nabla_{\mu}B^{\mu}_{\nu}+\frac{2(4-z)}{z+2}\frac{\nabla_{\mu}\varphi}{\varphi}
B^{\mu}_{\nu}+\frac{2(z\lambda-1)}{z+2}\frac{\nabla_{\nu}\varphi}{\varphi}
B-\lambda\nabla_{\nu}B=0,\label{Nieq}\\
&&\hspace*{-3em}\delta_{g^{\mu\nu}}S_{5{\rm D}};~~N\varphi^{2}E_{\mu\nu}^{(1)}+\beta_1\varphi^{-\frac{2(z-4)}{z+2}}E_{\mu\nu}^{(2)}
+\beta_1\lambda\varphi^{-\frac{2(z-4)}{z+2}}E_{\mu\nu}^{(3)}+N\beta_2\varphi^{2}E_{\mu\nu}^{(4)}=0,\label{gijeq}
\end{eqnarray}
where
\begin{eqnarray}
E_{\mu\nu}^{(1)}&=&R^{(4)}_{\mu\nu}-\frac{1}{2}g_{\mu\nu}R_{(4)}+\frac{2(2z+1)}{z+2}\frac{\nabla_{\gamma}N}{N}
\frac{\nabla^{\gamma}\varphi}{\varphi}g_{\mu\nu}+\frac{2(z^2-2z-2)}{(z+2)^2}
\frac{\nabla_{\gamma}\varphi}{\varphi}\frac{\nabla^{\gamma}\varphi}{\varphi}g_{\mu\nu}\nn\\
&&+\frac{4(1-z)}{z+2}\frac{\nabla_{(\mu}N}{N}\frac{\nabla_{\nu)}\varphi}{\varphi}
-\frac{2(z^2-8z-8)}{(z+2)^2}\frac{\nabla_{\mu}\varphi}{\varphi}\frac{\nabla_{\nu}\varphi}{\varphi}
+g_{\mu\nu}\frac{\nabla_{\gamma}\nabla^{\gamma}N}{N}\nn\\
&&+2g_{\mu\nu}\frac{\nabla_{\gamma}\nabla^{\gamma}\varphi}{\varphi}-\frac{\nabla_{\mu}\nabla_{\nu}N}{N}
-2\frac{\nabla_{\mu}\nabla_{\nu}\varphi}{\varphi}+\frac{1}{2}g_{\mu\nu}\varphi^{-2}V(\varphi),\nonumber\\
E_{\mu\nu}^{(2)}&=&-2N_{(\mu}\nabla_{|\gamma|}B_{\nu)}^{\gamma}
+2\nabla_{(\nu}N^{\rho}B_{\mu)\rho}+N_{\rho}\nabla^{\rho}B_{\mu\nu}+2NB_{\rho\mu}B_{\nu}^{\rho}
+\frac{N}{2}B_{\rho\sigma}B^{\rho\sigma}g_{\mu\nu}-NBB_{\mu\nu}\nn\\
&&-\partial_{y}B_{\mu\nu} +\left(z-2\right)\theta
B_{\mu\nu}+\frac{4(z-4)}{z+2}N_{(\mu}B_{\nu)}^{\rho}\frac{\nabla_{\rho}\varphi}{\varphi}
+\frac{4}{z+2}B\frac{N_{(\mu}\nabla_{\nu)}\varphi}{\varphi},\nonumber\\
E_{\mu\nu}^{(3)}&=&-N\frac{B^2}{2}g_{\mu\nu}+NB^2g_{\mu\nu}+2\nabla_{(\mu}B
N_{\nu)}-\frac{4z}{z+2}B\frac{N_{(\mu}\nabla_{\nu)}\varphi}{\varphi}-z\theta Bg_{\mu\nu}\nn\\
&&-\nabla_{\rho}BN^{\rho}g_{\mu\nu}+\partial_{y}Bg_{\mu\nu},\nonumber\\
E_{\mu\nu}^{(4)}&=&A_{\mu}A_{\nu}-\frac{1}{2}g_{\mu\nu}A_{\gamma}A^{\gamma},
\nonumber
\end{eqnarray}
and
\begin{eqnarray}
&&\hspace*{-4em}\delta_{\varphi} S_{5{\rm D}};~~2\varphi R_{(4)}-\frac{48(z+1)}{(z+2)^2}\nabla_{\gamma}{\nabla^{\gamma}}\varphi
-\frac{12}{(z+2)N}\varphi\nabla_{\rho}\nabla^{\rho}N
-\frac{48(z+1)}{(z+2)^2N}\nabla_{\rho}\varphi\nabla^{\rho}N
\nn\\
&&\hspace*{-3em}+\frac{2\beta_1}{z+2}\varphi^{-\frac{3(z-2)}{z+2}}\Big[\{2-\lambda(z+4)\}
B^2-\frac{2}{N}(1-4\lambda)zB\theta
-\frac{2}{N}\nabla_{\rho}BN^{\rho}+(z-4)B_{\mu\nu}^2\nn\\
&&\hspace*{-3em}+2(1-4\lambda)\frac{1}{N}\partial_{y}B\Big]-2\beta_2\varphi\Big[A_{\mu}A^{\mu}-\frac{4}{z+2}\frac{\nabla_{\mu}N}{N}A^{\mu}
+\frac{2z}{z+2}\nabla_{\mu}A^{\mu}\Big]-\frac{d
V(\varphi)}{d\varphi}=0, \label{phieq0}
\end{eqnarray}
where $\theta=2(\partial_y{\varphi}-\nabla_{\rho}\varphi
N^{\rho})/\{(z+2)\varphi\}$.

\subsection{Exact solutions}

In order to find ground state  solutions for the
Eqs.(\ref{Neq})$\sim$(\ref{phieq0}) with $y$-dependence, we work in the  comoving gauge  with $N_{\mu}=0$,   where the five dimensional metric tensor
is given by
\begin{eqnarray}\label{met}
ds^2&=&g_{\mu\nu}dx^\mu dx^\nu +N^2dy^2.
\end{eqnarray}
Let us   consider the
ansatz\begin{eqnarray}\label{ansat}
g_{\mu\nu}=a^2(y)\eta_{\mu\nu},~~~N=N(y),~~~\varphi=\varphi(y),
\end{eqnarray}
which describes a warped spacetime \cite{raych} with  $4D$ Minkowski vacuum ($R^{(4)}_{\mu\nu}=0$). We include a matter term $S_m$ in the action (\ref{conformalR}) at this stage and substitution of the above ansatz (\ref{ansat}) into the Eqs.(\ref{Neq})$\sim$(\ref{phieq0}) (plus matter) yields the following equations:
\begin{eqnarray}
\hspace*{-1em}(1-4\lambda)\beta_1{\cal H}^2&=&-2(1-4\lambda)\beta_1{\cal H}\theta-(1-4\lambda)\beta_1\theta^2
+\frac{N^2}{4}\varphi^{z-4}V+\frac{N^2}{4}\varphi^{\frac{2(z-4)}{z+2}}{\cal T}_{1},\label{eqH}\\
\hspace*{-1em}(1-4\lambda)\beta_1\partial_{y}{\cal H}&=&(1-4\lambda)\beta_1\Big[z\theta^2+z{\cal H}\theta-\partial_y \theta+\frac{\partial_y N}{N}({\cal H}+\theta)\Big]-\frac{N^2}{2}\varphi^{\frac{2(z-4)}{z+2}}({\cal T}_1+{\cal T}_2),\nn\\
&&\label{eqHp}\\
0&=&(1-4\lambda)\beta_1\Big[(z+4){\cal H}^2-(z-4)\theta^2+8{\cal H}\theta+2(\partial_y{\cal H}+\partial_y\theta)\nonumber\\
&&\hspace*{11em}-\frac{2\partial_y N}{N}({\cal H}+\theta)\Big]-\frac{N^2}{4}\varphi^{\frac{2(z-3)}{z+2}}\frac{dV}{d\varphi},
\label{eqphi}
\end{eqnarray}
where quantities ${\cal H}$ and ${\cal T}_{1,2}$ are given by
\begin{eqnarray}
{\cal
H}=\frac{\partial_y a}{a},~~~~~{\cal T}_1=-\frac{1}{M_{*}^{3}\sqrt{g_{(4)}}}\frac{\delta S_M}{\delta
N},~~~~~{\cal T}_2g_{\mu\nu}=-\frac{2}{N M_{*}^{3}\sqrt{g_{(4)}}}\frac{\delta S_M}{\delta
g^{\mu\nu}}.
\end{eqnarray}
 After some manipulations of the
Eqs.(\ref{eqH})$\sim$(\ref{eqphi}), one can combine the above
equations into a single equation:
\begin{eqnarray}\label{condi}
{\cal T}~=~(z+4)V-\frac{z+2}{2}\varphi \frac{dV}{d\varphi},
\end{eqnarray}
where a quantity ${\cal T}$ is defined by ${\cal T}\equiv-z{\cal
T}_1+4{\cal T}_2$. We note that ${\cal T}$ is related to the trace of
the energy momentum tensor for the matter $S_{m}$. In an isotropic
case with $z=1$, ${\cal T}$ corresponds to the trace $T_{A}^{A}$ for the
energy momentum tensor $T_{AB}$ of matter in 5D. Note also that for the anisotropic
invariant action (\ref{conformalR}) with the potential (\ref{pote}),
the $r.h.s.$ of the Eq.(\ref{condi}) yields zero so in this case, the quantity
${\cal T}$ should satisfy the following condition
\begin{eqnarray}\label{t1t2}
{\cal T}=0~\to~{\cal T}_2=\frac{z}{4}{\cal T}_1.\label{confcond}
\end{eqnarray}
Alternatively, the above condition (\ref{t1t2}) can be obtained,
after some tedious manipulations with
Eqs.(\ref{Neq})$\sim$(\ref{phieq0}), by checking $\delta S_{5{\rm
D}}=0$ for the infinitesimal ($\omega\ll1$) transformations
(\ref{trans}): \bea \delta N= z\omega N,~~~~\delta N^{\mu}=0,~~~~
\delta g^{\mu\nu}= -2\omega g^{\mu\nu},~~~~\delta \varphi= -\frac{z+2}{2}\omega
\varphi,\nonumber \eea \bea &&\hspace*{-5em}\delta S_{5{\rm D}}~=~
\frac{\delta S_{5{\rm D}}}{\delta N}\delta N+ \frac{\delta S_{5{\rm
D}}}{\delta g^{\mu\nu}}\delta g^{\mu\nu}+ \frac{\delta S_{5{\rm
D}}}{\delta N^{\mu}}\delta N^{\mu}+
\frac{\delta S_{5{\rm D}}}{\delta \varphi}\delta \varphi~=~0\nn\\
&&\nonumber\\ &&\hspace*{3.5em}\rightarrow~{\cal
T}_2=\frac{z}{4}{\cal T}_1 \eea
The above equation says that only matter fields satisfying this condition can couple. In other words, Eq. (\ref{confcond}) is the anisotropic conformal condition that should be satisfied by the matter sector.

Note that  the potential (\ref{fracpote}) satisfies  Eq. (\ref{confcond}) and we present solutions with $(i)$ $V_0=0$ and $(ii)$ $V_0\neq 0$ :
 Explicitly, one  finds
the solutions as follows:
\begin{center}
$(i)~V_0=0$\\[3mm]
$g_{\mu\nu}=e^{-2\zeta^2 y^2}\eta_{\mu\nu},
~~~~~N= N_0e^{-{b\zeta^2}y^2},~~~~~\varphi=\varphi_0
e^{\frac{z+2}{2}\zeta^2y^2};$\\[6mm]
$(ii)~V_0=16(1-4\lambda)\frac{\zeta^4
y_0^2}{N_0^2}\varphi_0^{-\frac{4z}{z+2}}$\\[3mm]
$g_{\mu\nu}=e^{-2\zeta^2(y-y_0)^2}
\eta_{\mu\nu},~~~~~N= N_0e^{-z{\zeta^2}y^2},~~~~~\varphi=\varphi_0
e^{\frac{z+2}{2}\zeta^2y^2},$
\end{center}
where $\zeta,~b,~y_0,~N_0,~\varphi_0$ are some constants.
Both of these solutions describe warped spacetime with Gaussian warp factor which has been discussed in a number of papers \cite{Flanagan:2001dy,Alexandre:2008ja,Quiros:2012bh,Moon:2016pzj}.
But we will not go further along this direction. Instead we proceed in the next section to investigate scalar perturbations and vacuum instability of case $(i)$.

\subsection{Scalar perturbations in the quadratic action}

It is known that an explicit breaking of the five dimensional diffeomorphism invariance leads to a pathological behavior of the scalar graviton \cite{Papantonopoulos:2011xa}, which includes four time derivatives or ghost-like kinetic term of the graviton. Thus, it is a very important issue to check if such a scalar graviton problem does exist in the model (\ref{conformalR}), which explicitly breaks the 5D diffeomorphism invariance.

In order to investigate the behavior of the scalar mode, we consider the following scalar perturbations of the metric and the scalar field for
the Minkowski background with $V_0=0$
\begin{eqnarray}
&&~N=e^{\alpha},~~~~~N_{\mu}=\partial_{\mu}\beta,\nn\\
&&g_{\mu\nu}=e^{2\xi}\eta_{\mu\nu},
~~\varphi=\varphi_0+\tilde\varphi,\label{pphi}
\end{eqnarray}
where $\alpha,~\beta,~\xi,~\tilde\varphi$ are function of
$(x^{\mu},~y)$ and $\varphi_0$ is a constant. We notice that the above ones (\ref{pphi}) correspond to the perturbations around the 5D Minkowski background solution $(i)$, obtained when choosing $\zeta=0$, $N_0=1$. Substituting the perturbations (\ref{pphi}) into the action (\ref{conformalR}) and after a lengthy computation, we arrive at  the following
quadratic action:
\begin{eqnarray}
&&\hspace*{-1.5em}S^{(2)}=\int \,dy
d^4x\Bigg[-6\varphi_0^{2}\Big\{(\partial_{\mu}\xi)^2+\alpha\Box^{(4)}\xi
+2\xi\Box^{(4)}\xi+2\varphi_0^{-1}\tilde\varphi\Box^{(4)}\xi\Big\}\nn\\
&&\hspace*{6em}
-\frac{12}{z+2}\varphi_0\Big\{2\partial_{\mu}\xi\partial^{\mu}\tilde\varphi+\alpha\Box^{(4)}\tilde\varphi+2\xi\Box^{(4)}\tilde\varphi
+\frac{2(z+1)}{z+2}\varphi_0^{-1}\tilde\varphi\Box^{(4)}\tilde\varphi\Big\}\nn\\
&&\hspace*{4.5em}+\beta_1\varphi_0^{-\frac{2(z-4)}{z+2}}\Big\{4(1-4\lambda)(\partial_{y}\xi)^2-2(1-4\lambda)\partial_y\xi\Box^{(4)}\beta+(1-\lambda)(\Box^{(4)}\beta)^2
\nn\\
&&\hspace{9.5em}+\frac{16(1-4\lambda)}{z+2}\varphi_0^{-1}\partial_{y}\xi\partial_{y}\tilde\varphi
-\frac{4(1-4\lambda)}{z+2}\varphi_0^{-1}\Box^{(4)}\beta\partial_{y}\tilde\varphi
\nn\\
&&\hspace*{8em}+\frac{16(1-4\lambda)}{(z+2)^2}\varphi_0^{-2}(\partial_{y}\tilde\varphi)^2\Big\}
+\beta_2\varphi_0^{2}\Big(\partial_{\mu}\alpha+\frac{2z}{z+2}\varphi_0^{-1}\partial_{\mu}\tilde\varphi\Big)^2\Bigg].
\label{qaction1}
\end{eqnarray}
Here, $\Box^{(4)}$ denotes the four dimensional d'Alembert operator, defined by $\Box^{(4)}\equiv-\partial_{t}^2+\partial_{{\bf x}}^2$. The action (\ref{qaction1}) includes a potentially dangerous term with the four time derivatives, given by $(\Box^{(4)}\beta)^2$ in the third line  and this raises vacuum instability problem. We check whether this term can be removed as a consequences of equations of motion for the fluctuations.

Varying the above action (\ref{qaction1}) with respect to $\alpha$ and $\beta$, one obtains the equations of motion as
\begin{eqnarray}
&&3\Box^{(4)}\xi+\frac{2(3+\beta_2 z)}{z+2}\varphi_0^{-1}\Box^{(4)}\tilde\varphi+\beta_2\Box^{(4)}\alpha=0,\label{phieq}\\
&&(1-\lambda)\Box^{(4)}\beta=(1-4\lambda)\left(\partial_{y}\xi
+\frac{2}{z+2}\varphi_0^{-1}\partial_y\tilde\varphi\right)\label{Beq}.
\end{eqnarray}
Note that for the case of $\beta_2=0$, $\alpha$ behaves like a Lagrange multiplier. In this case, the perturbations $\varphi$, $\beta$ can be completely eliminated due to the Eqs. (\ref{phieq}), (\ref{Beq}) in favor of the scalar graviton mode $\xi$. It turns out that in this case, there is no dynamical scalar graviton in the quadratic action (\ref{qaction1}).

On the other hand, for $\beta_2\neq0$ one can replace $\alpha$, $\Box^{(4)}\beta$ with $\xi$, $\tilde\varphi$ in the action (\ref{qaction1}) by using the Eqs. (\ref{phieq}), (\ref{Beq}). It is found that after taking integration by parts, the quadratic action (\ref{qaction1}) can be written as
\begin{eqnarray}
S^{(2)}&=&\int \,dy d^4x\Big\{A_{1\xi}
(\partial^{(4)}\xi)^2+A_{1\varphi}
(\partial^{(4)}\varphi)^2+A_{2\xi}
(\partial_y\xi)^2+A_{2\varphi}
(\partial_y\varphi)^2\nonumber\\
&&\hspace*{10em}+A_{1\xi\varphi}
\partial^{(4)}\xi\partial^{(4)}\varphi+A_{2\xi\varphi}
\partial_y\xi\partial_y\varphi \Big\},\label{quad2}
\end{eqnarray}
where
\begin{eqnarray}
&&\hspace*{-2.5em}A_{1\xi}=\frac{6\varphi_0^{2}(\beta_2-\frac{3}{2})}{\beta_2},
~~A_{1\varphi}=\frac{24(\beta_2-\frac{3}{2})}{(z+2)^2\beta_2},
~~A_{1\xi\varphi}=\frac{24\varphi_0(\beta_2-\frac{3}{2})}{(z+2)\beta_2},\nonumber\\
&&\hspace*{-2.5em}A_{2\xi}=3\varphi_0^{-\frac{2(z-4)}{z+2}}\frac{(4\lambda-1)\beta_1}{\lambda-1},
~~A_{2\varphi}=\frac{12\varphi_0^{-\frac{4(z-1)}{z+2}}}{(\lambda-1)(z+2)^2},
~~A_{2\xi\varphi}=\frac{12\varphi_0^{-\frac{3(z-2)}{z+2}}}{(\lambda-1)(z+2)}.
\end{eqnarray}
It should be pointed out that the resulting quadratic action (\ref{quad2}) does not include the fourth order derivative terms and also, the problems of the ghost/instability can be cured for
\begin{eqnarray}\label{range}
0<\beta_2<\frac{3}{2},~~~~~~~~\beta_1\left\{\begin{array}{ll}
>0~~~(1/4<\lambda<1) \label{soo1}\\
<0~~~(\lambda<1/4~{\rm or}~1<\lambda) \label{soo2}\end{array}\right.,
\end{eqnarray}
which corresponds to $A_{1\xi,1\varphi}<0,~A_{2\xi,2\varphi}<0$, respectively. Also, it can be checked that for $\lambda=1$ or $\lambda=1/4$, substitution of (\ref{phieq}) and (\ref{Beq}) into (\ref{qaction1}) leads to
\begin{eqnarray}
S^{(2)}&=&\int \,dy d^4x\Big\{A_{1\xi}
(\partial^{(4)}\xi)^2+A_{1\varphi}
(\partial^{(4)}\varphi)^2+A_{1\xi\varphi}
\partial^{(4)}\xi\partial^{(4)}\varphi\Big\},\label{quad3}
\end{eqnarray}
which implies that there are no the ghost/instability problems for $0<\beta_2<3/2$.

Consequently, we remark that the above result (\ref{range}) is in contrast to that in \cite{Papantonopoulos:2011xa} without the Weyl scalar mode, where there remains ghost scalar mode in the spectrum: either from the quadratic kinetic term, or from the higher time derivative terms. It is to be noticed  that the Weyl scalar mode $\tilde\varphi$ in the quadratic action (\ref{qaction1}) plays an essential role of taming the scalar graviton $\xi$ and fourth order derivative terms, as in Eqs.(\ref{qaction1})$\sim$(\ref{Beq}), thus curing  the pathologies of other modes at least  in the quadratic perturbations (\ref{quad2}). It remains to be checked whether  such nice behavior persists at the cubic order\footnote{It was shown in \cite{Moon:2011xi,Moon:2013ska} that the 4D anisotropic action with scalar field does not show any pathological behaviors at the quadratic as well as cubic order in the perturbative action.}.

\section{Conclusion and discussions}

In this paper, we considered the higher dimensional anisotropic invariant gravity ($z-$Weyl gravity) in which the four dimensional spacetime and extra dimensions are not treated on an equal footing. In order to implement the anisotropy, we considered higher dimensional  ADM decomposition and impose
the foliation preserving diffeomorphism where the foliation is adapted along the extra dimensions, thus keeping  the general covariance for the four dimensional spacetime. The resulting gravity is extended to the conformally invariant case with an extra (Weyl) scalar field as usual, but with a new ingredient of a parameter $z$ which describes the degree of anisotropy of conformal transformation between the spacetime and extra dimensional metrics. Effective four-dimensional gravity reduces to a  scalar-tensor theory
coupled with nonlinear sigma model. We have performed scalar perturbations in the quadratic action in 4+1 dimensions, and found a range of parameter space which does not lead to four time derivatives or ghost-like kinetic term of the scalar  graviton. Extension of the analysis to the general $4+D$ dimensional case remains to be done.

The anisotropic conformal invariance puts no restriction a priori on the value of the conformal weight $z$, at least at the classical level and this may lead to some new aspects  hidden in the anisotropic theory.
For example, in $5D,$ the isotropic potential yields a fractional power $V\sim \phi^{\frac{10}{3}}$\cite{Moon:2016pzj}.
Such a potential renders  a perturbative
 approach inaccessible, and could be  plagued  with possible discontinuities of the potential. On the other hand, in the anisotropic case, any even   power is allowed. This has the potential to provide a single framework for the descriptions of   both positive and negative powers of potentials

In fact, the conformal weight is a dynamical   exponent: One can easily check the global scaling symmetry of the action (\ref{confact1})
under the following transformations
 \begin{equation}
 x_\mu\rightarrow b x_\mu,~ y_m\rightarrow b^{z}y_m, ~N_m^\mu\rightarrow b^{1-z}N_m^\mu, ~\phi\rightarrow b^{-v(z)}\phi. \label{scaling1}
 \end{equation}
Horava-Lifshitz gravity exhibits dynamical scaling with $z = 3$ in the UV, flowing to the relativistic value
$z = 1$ in the IR where Lorentz symmetry is restored.
Since there is no symmetry that protects $z$, quantum corrections would modify the scaling laws, producing non-integer $z$'s in general.
This leads to consider some cases where an exotic values of $z$ can be considered. In Ref.\cite{Porto:2009xj}, dynamical critical behavior of gravity in the extreme ultra-infrared
 sector and a mechanism to relax the cosmological constant was speculated and it is argued that  $z\sim 20-30$ would produce the correct factor $\sim 10^{-120}$ to address the cosmological constant problem. In our case,
perhaps the most interesting aspect arises when $z$ assumes a negative value close to $-2.$ If we consider the cosmological constant given by (\ref{cc1}) with $D=1$,  we have (with $V_0\sim {\cal O}(1))$
\begin{equation}
\Lambda\sim M_*^2 \varphi_0^{\frac{4}{z+2}},
\end{equation}
Note that the detailed value of $\Lambda$ does not depend on reduction scheme (or brane world scenario) like the size of the extra dimension, but only on the scale  $M_*$ and $z$. Then the essential point is that when $z$ is very close to $-2$, the $\varphi_0$ part can produce a very small number. For example, with $\varphi_0\sim 10$ and $z=-2.04~(N=-49 $ in Eq, (\ref{parity})), this factor can produce a  number like $\sim 10^{-100}$.  More fine-tuning would be necessary
in order to produce the correct factor for the cosmological constant problem, but such tuning can be compared with the conventional
one which requires a fine-tuning of the order
$10^{-120}$ for the cosmological constant.

The negative critical exponent is very rare in physical situations \cite{Hertz:1976zz}, but that does not necessarily mean that it is forbidden at a fundamental level.
 Such a negative value  would enforce a bizarre scaling properties of extra dimensions, scaling like  positive powers of  mass dimension\footnote{In the so-called space-time-matter theory \cite{Overduin:1998pn}, the extra dimension is treated as an explicit mass dimension with $y=G_Nm/c^2$.}.
It also gives unconventional uncertainty relation for  a negative value  of $z.$
Interpreting the scaling law (\ref{scaling1}) in energy-momentum space, we obtain  scaling relations like $E\sim p_x\sim (p_y)^{\frac{1}{z}}$. As the energy  approaches UV region, the spatial $x-$momentum also increases, but the $y$-momentum along the extra dimensions decreases, and vice versa for a negative value of $z$. Another way of saying this is that if we probe a short distance behavior in $x$-space with high energy, we are at the same time looking at the long-distance properties in the extra dimensions. This type of behavior seems to be counter intuitive from the ordinary quantum mechanical point of view, but is reminiscent in spirit of the T-duality \cite{Zwiebach:2004tj}
in string theory which does not distinguish the large scale from the smaller one.

It is to be  remarked that  the idea of negative dynamical exponent for the extra dimension and approaching the cosmological constant problem based upon it remains an open possibility.  One could perform a cosmological analysis and compare with observational data in order to actually test the value of
$z.$ This work is in progress and will be reported elsewhere.


\section*{Acknowledgments}
We would like to thank Soo-Jong Rey and Sang-Jin Sin for useful
discussions. This work was supported by Basic Science Research
Program through the National Research Foundation of Korea (NRF)
funded by the Ministry of Education (Grant No. 2015R1D1A1A01056572).

\end{document}